\documentclass[twocolumns]{aa}
\usepackage{graphicx}
\usepackage{txfonts}

\begin{document}
\title{Q2237+0305 in X-rays: spectra and variability with XMM-Newton}

\author{E.V.Fedorova \inst{1}, V.I.Zhdanov \inst{1}, C.Vignali \inst{2}, G.G.C.Palumbo \inst{2}}

\offprints{E.V.Fedorova.}

\institute{Astronomical Observatory of Kyiv Shevchenko
University, Observatorna St. 3, Kiev, Ukraine, UA- 04053 \\
              \email{elena\_f@mail.univ.kiev.ua, zhdanov@observ.univ.kiev.ua}
         \and
             Dipartimento di Astronomia, Universit\`{a} di Bologna, via Ranzani 1, 40127, Bologna, Italy\\
             \email{cristian.vignali@unibo.it, giorgio.palumbo@unibo.it}
             \thanks{Based on observations
obtained with XMM-Newton, an ESA science mission with instruments
and contributions directly funded by ESA Member States and NASA.}
             }


  \titlerunning{2237+0305 in X-rays}

\authorrunning{Fedorova et al}

  \abstract
   {}
   {X-ray observations of gravitationally lensed quasars may allow
   us to probe the inner structure of the central engine of a quasar.
   Observations of Q2237+0305 (Einstein Cross) in X-rays may be used to
   constrain the inner structure of the X-ray emitting source.}
   {Here we analyze the XMM-$\it{Newton}$ observation of the quasar in the gravitational lens system Q2237+0305 taken
   during 2002. Combined spectra of the four images of the quasar in this system
   were extracted and modelled with a power-law model. Statistical analysis
   was used to test the variability of the total flux.}
   {The total X-ray flux from all the images of this quadruple gravitational lens system is
$6\cdot10^{-13}$ erg/(cm$^2\cdot$sec) in the range 0.2--10~keV,
showing no significant X-ray spectral variability during
almost 42~ks of the observation time. Fitting of the cleaned
source spectrum yields a photon power-law index of
$\Gamma=1.82_{ - 0.08}^{ + 0.07}$. The X-ray lightcurves obtained
after background subtraction are compatible with the hypothesis of
a stationary flux from the source.}
   {}

   \keywords {gravitational lensing -- X-rays: general -- quasars: general}

   \maketitle
%

\section{Introduction}

Since its discovery, the gravitational lens system (GLS) Q2237+0305 (Einstein Cross; 
(Huchra et al., \cite{huchra}) has attracted much 
attention as a unique laboratory to study gravitational
lensing effects. This GLS consists of a quadruply-imaged quasar
and a lensing galaxy that is the nearest among all known
gravitational lens systems (the quasar redshift is 
$z_Q=1.695$, while the lens redshift is $z_G=0.0395$). Due to its very
convenient configuration, it is one of the best investigated GLSs.
It has been continuously monitored by a number of groups from 
1992--2005 (Rix et al., \cite{rix}; Falco et al., \cite{vla};
Oestensen et al., \cite{not}; Blanton et al., \cite{hubble};
Bliokh et al., \cite{maidanak}; Nadeau et al., \cite{monica};
Wozniak et al., \cite{ogle}; Alcalde et al., \cite{glitp}; Schmidt
et al., \cite{apo}). Significant microlensing-induced brightness
peaks on lightcurves of quasar images were detected in this system
(see, e.g., Wozniak et al., \cite{ogle}; Alcalde et al.
\cite{glitp}; Moreau et al., \cite{moreau}). The OGLE
group is continuing observations of the Einstein Cross and 
the database $http://bulge.princeton.edu/\sim ogle$ is being
constantly renewed.

Observations of the Einstein Cross in X-rays are likely to
provide important data that can be used to constrain the source
inner structure. It is hoped that the detection of a
variable X-ray flux from this system will lead to estimates of the
relative time-delays $\Delta\tau$ between the images.

The X-ray emission of Q2237+0305 was first detected during {\it ROSAT} 
observations in 1997 (Wambsganss et al., \cite{rosat}). From the
analysis of Wambsganss et al. (\cite{rosat}), a 0.1-2.4 keV count
rate of 0.006 count/sec was obtained and, assuming a $\Gamma=1.5$
power-law model and Galactic absorption (hydrogen column density
$N_{H}=5.5\cdot 10^{20}$~cm$^{-2}$; Dickey \& Lockman, \cite{HI}),
a flux of $2.2\cdot 10^{-13}$ erg/(cm$^{2}\cdot$ s) was derived.
The {\it ROSAT} observation of Q2237+0305 did not
show any significant flux variability, and the spatial resolution
of the {\it ROSAT}/HRI detector is not sufficient to resolve the
different quasar images.

After {\it ROSAT}, GLS Q2237+0305 was observed several times with the Advanced CCD
Imaging spectrometer onboard the $\it{Chandra\ X-ray\
Observatory}$. Results from a 30~ks and a 
10~ks  observation of Q2237+0305, carried out on September 2000 and on
December 2001, respectively, were published by Dai et al.
(\cite{dai}). For the former of these $\it{Chandra}$ observations, 
the X-ray flux was $4.6 \cdot 10^{-13}$ erg/(cm$^{2}\cdot$s) in
the energy range 0.4--8.0~keV, and the lensed luminosity was $1.0
\cdot 10^{46}$ erg/s; for the latter observation, the flux was $3.7
\cdot 10^{-13}$ erg/(cm$^{2}\cdot$ s) and the lensed luminosity
$8.3 \cdot 10^{45}$ erg/s (in the same energy range).

An important point concerns the variability of the X-ray signal
from the source that carries useful information about the
innermost source structure. In the case of Q2237+0305, the
detection of X-ray variability is especially important because it
may lead to a direct estimate of the gravitational time-delays
that are a significant characteristic of lensed systems. The model
prediction of $\Delta\tau$ between different Einstein Cross images
is of the order of hours (Schmidt et al. \cite{schmidt}), whereas
the optical data may provide time-delay values that are accurate
to only 1-2 days (Vakulik et al., \cite{vakulik}). The only
observational estimate of the shortest time-delay is obtained by
cross-correlating the X-ray lightcurves from different images of
Q2237+0305 (Dai et al., \cite{dai}) of the $\it{Chandra}$
observations: here all four X-ray images of this quasar have been
resolved and sufficient variability has been reported (the difference between total
fluxes during the first and the second $\it{Chandra}$ observations
is close to 20$\%$ of the averaged value). The
latter observation enabled Dai and collaborators to determine the
time delay of $\Delta\tau_{BA}= 2.7_{-0.9}^{+0.5}$~hours between
two of the four GLS images. The delays of the other images of
Einstein Cross are still unknown.

 The XMM-$\it{Newton}$
receivers cannot separate different images of the Einstein Cross.
Nevertheless, one may hope to obtain some useful information about
the source from the autocorrelation function of the total
X-ray flux from all the images. The applicability of this method
depends on the presence of sufficient flux variability and may be
complicated by the presence of possible microlensing.

In the present paper we analyze the XMM-$\it{Newton}$ observation
of the GLS Q2237+0305 taken in 2002. We describe the data
reduction and present the parameters of the X-ray spectra obtained
by EPIC cameras and RGS spectrometers onboard 
XMM-$\it{Newton}$. Also, we present the results on variability in the range 0.2--10~keV. Then
we discuss an estimate of the source inhomogeneity scale in view
of high-magnification microlensing events in this GLS.

\begin{figure}
 \resizebox{\hsize } {!} { \includegraphics{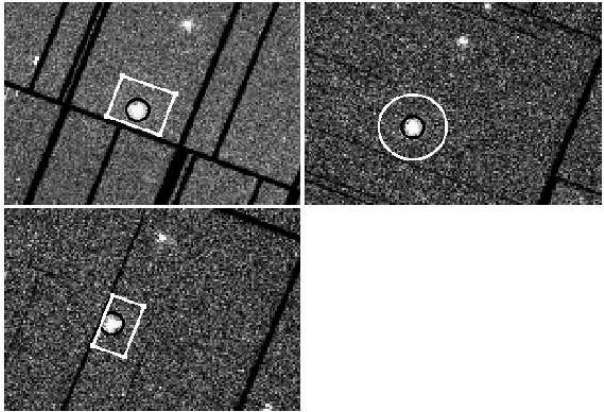} }
 \caption{EPIC pn (upper left), MOS1 (upper right) and  MOS2 (below) images of the Q2237+0305.
 Black circles define areas collecting source signal, while white-bordered areas show
 the regions chosen for estimates of the background.}
 \label{fig1}
 \end{figure}

\section{Data reduction}

GLS Q2237+0305 was observed with XMM-$\it{Newton}$ for an exposure
time of 43.9~ks in 28 May 2002 (PI: Michael Watson; XMM-{\it Newton} 
public archive). Observation ID, date of observation, total
source counts and count rates in the extraction regions can be
found in Table \ref {tab1}.

\begin{table}
\begin{tabular}{|l|c|c|c|} \hline
{dataset ID } & & {0110960101}\\\hline {observation date} &  &
{28.05.2002} \\\hline
& MOS & 42.3\\
{exposure time, (ks)} & pn & 39.6 \\
& RGS & 42.8 \\\hline
& MOS1 & 0.08\\
& MOS2 & 0.06\\
{count rate, (count/s)}& pn &0.29\\
& RGS1 & 0.11\\
& RGS2 & 0.10\\\hline
& MOS1 & 3170\\
& MOS2 & 2542\\
{total source counts}& pn &11446\\
& RGS1 & 4707\\
& RGS2 & 4279\\\hline
\end{tabular}
\vspace{0.05in} \caption{XMM-Newton observation log for
Q2237+0305. Here and throughout the paper the name ``MOS" is used
for the MOS1 and MOS2 together.} \label{tab1}
\end{table}

The European Photon Imaging Cameras (EPIC) MOS1, MOS2 and pn
detectors were operated in full frame mode for the whole time of
the observation. EPIC data and the data from the Reflection
Grating Spectrometers (RGS) were processed with the
XMM-$\it{Newton}$ Standard Analysis System ({\sc sas}, public release
version 6.5.0).\footnote{{\sc sas} software and its description can be
found at http://xmm.esac.esa.int/sas/.} After filtering of
bad-flagged data, the effective exposure times were: 39.6~ks for
pn-camera, 42.3~ks for MOS-cameras and 42.8~ks for
RGS-spectrometers. The EPIC maps of Q2237+0305 are shown in
Fig.\ref{fig1}.

\begin{figure}
 \resizebox{\hsize } {!} { \includegraphics{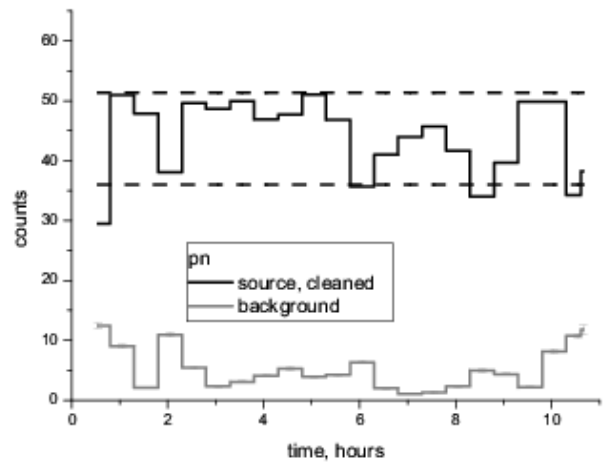} }
 \caption{The full band (0.2-10~keV) EPIC pn  lightcurves
 for the source (background-subtracted) and the background. The dashed straight lines show the
  $1\sigma$ limits of the averaged value of the flux during the observation time. The time bin size is 1.8~ks.}
 \label{fig2}
 \end{figure}

\begin{figure}
 \resizebox{\hsize } {!} { \includegraphics{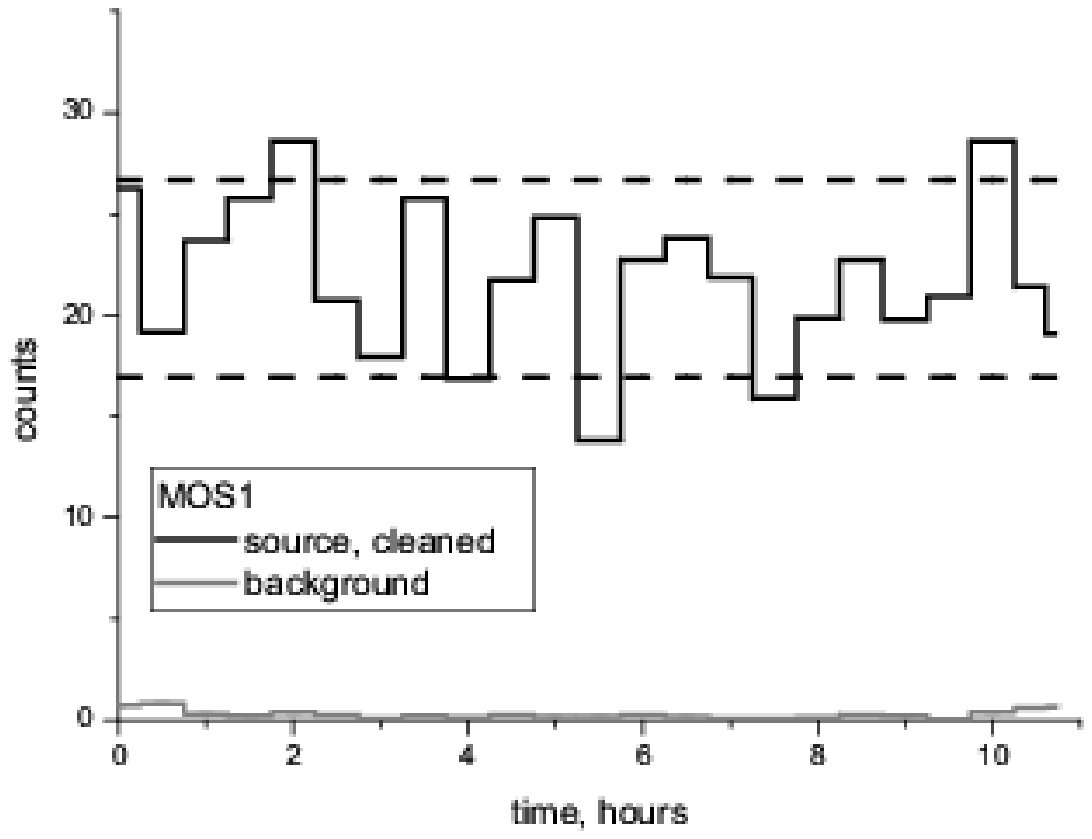} }
 \caption{The same for EPIC MOS1 as in Fig. \ref{fig2}.}
 \label{fig3}
 \end{figure}

\begin{figure}
 \resizebox{\hsize } {!} { \includegraphics{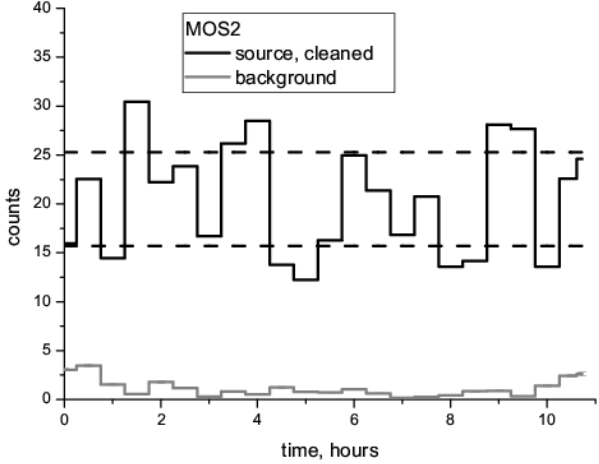} }
 \caption{The same for EPIC MOS2 as in Fig. \ref{fig2}.}
 \label{fig4}
 \end{figure}

The energy range of EPIC cameras is 0.2--10~keV, and the RGS
energy range is 0.3--2.8~keV. To obtain the cleaned lightcurves,
we extracted full count rates from circular areas with different
radii around the source (see Fig.~\ref{fig1}). To estimate the
background effects we also used areas of different sizes near the
source on the same CCD plates. Flux variations from different
background areas agree with the corresponding estimated Poisson
variance. These variations have been taken into account in the
final error estimates of the resulting net signal. The background
counts were subtracted (with corresponding factors taking into
account different background areas) from the total counts from the
source regions to obtain the cleaned lightcurves. The best
signal-to-background ratios at low enough levels of Poisson
variance were obtained for the case of 3-5~arcsec radii of the
source areas. The background-subtracted lightcurves are shown in 
Figs.~\ref{fig2} -- \ref{fig4} together with the corresponding
background lightcurves.

\begin{figure}
 \resizebox{\hsize } {!} { \includegraphics{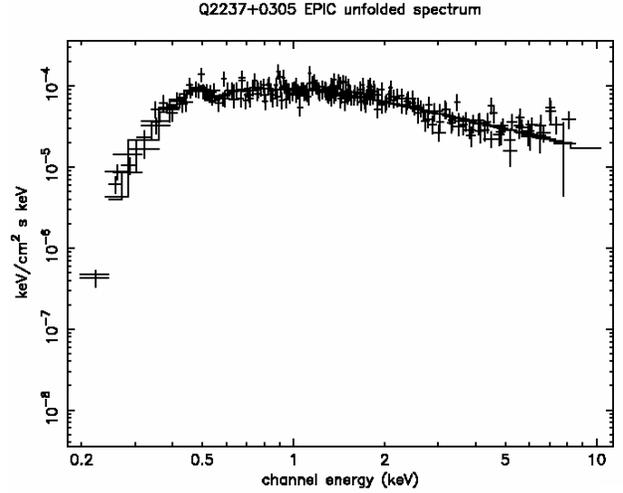} }
 \caption{Unfolded XMM-{\it Newton} EPIC (pn and MOS) spectra of Q2237+0305 
and power-law model. Here the 90\% error bars are shown.}
 \label{fig5}
 \end{figure}

EPIC spectra were obtained through standard {\sc sas} procedures:
evselect, arfgen, and rmfgen. For spectral filtering, the single
high-energy events were extracted in order to identify soft proton
flares ($E>10$~keV, for pattern zero only; the time-bin sizes were
100 s for the pn and 10 s for the MOS). Good time intervals were
defined using the standard {\sc sas} procedure tabgtigen (with rate
parameter 1.0 cts/s for the pn camera and 0.35 cts/s for the MOS)
and then the event files were corrected. The background
subtracted EPIC spectra are shown in Fig.~\ref{fig5}.

In the case of the lightcurves, we have not used tabgtigen, but
we have simply subtracted the background counts from the total ones 
in order to obtain a continuous time series.

RGS spectra were obtained using standard {\sc sas} tasks (rgsproc and
rgsrmfgen). RGS lightcurves were obtained and
background-subtracted similarly to the EPIC data (i.e., 
using an evselect procedure); they are characterized by larger errors than 
EPIC lightcurves. 
No significant absorption or emission lines were detected in these spectra. 

\section{Spectral fitting}
To fit the Q2237+0305 spectrum, as a first step we tried a
redshifted power-law model (zpowerlw)

\[
\rho (E) = KE^{ - \Gamma }
\]

\noindent (where $\Gamma$  is the photon index), with
photoelectric absorption (zphabs) by neutral material in the
lensing galaxy

\[
\rho (E) = Ke^{ - \frac{\sigma (E)N_H }{1 + z_G }}
\]
\noindent where ${\sigma (E)}$ is the photo-electric cross-section
and ${N_H }$ is hydrogen column density (see Table \ref{tab2}),
using {\sc xspec} v.~12.2.1. To calculate lensed luminosities, we used
these values for the cosmological parameters: Hubble constant
$H_{0}$=70 km/(s$\cdot$ Mpc), $\Lambda_{0}=0.7$ and
$\Omega_{M}=0.3$. We compare fluxes and lensed luminosities in
different rest-frame energy ranges obtained from our fit with
spectral fits from Dai et al. (\cite{dai}) and Wambsganss et al.
(\cite{rosat}, see Table \ref{tab3}). The row labeled MOS
in Table \ref{tab2} refers to the MOS1 and MOS2 spectra fitted
with the same model simultaneously, and the row labelled EPIC
represents the MOS1, MOS2 and pn spectra fitted with the same
model simultaneously. We obtained the values of $N_H$ for the
lensing galaxy taking into account the mentioned above frozen
value of the Galactic absorption (Dickey \& Lockman, \cite{HI}).
The obtained values of photon indices and hydrogen column
densities are in good agreement with the results of Dai et al.
(\cite{dai}), but our power-law indices are obviously higher than
the power-law index of 1.5, assumed by Wambsganss et al.
(\cite{rosat}), likely due to the wider spectral range.

\begin{table}
\begin{tabular}{|l|c|c|c|c|} \hline
 Camera  & {$N_{H}$,} & {$\Gamma$}& $\chi ^{2}$/d.o.f.& EW,\\
 & {$10^{20}$ cm$^{-2}$}& & & keV \\ \hline
 MOS1 &     $9_{ - 4}^{ + 5}$  & $1.86_{ - 0.16}^{ + 0.17}$ &  38.8/38 &$<1.6$\\ \hline
 MOS2 &     $8\pm$4            & $1.94_{ - 0.11}^{ + 0.14}$ &  38.6/41 &$<1.1$\\\hline
 pn   &     $5_{ - 2}^{ + 3}$  & $1.85_{ - 0.12}^{ + 0.15}$ &  93.0/84 &$<1.3$\\\hline
 MOS  &     $8\pm$3  &   $1.92\pm$0.10                      & 80.1/83  &\\\hline
 EPIC &     $7\pm$2  &   $1.88\pm$0.10                      & 183.5/168  &\\\hline
\end{tabular}
\vspace{0.05in} \caption{Absorbed power-law fitting of
XMM-$\it{Newton}$ Q2237+0305 spectra - hydrogen column density in
the lens galaxy (at a redshift $z_G=0.0395$), photon index and
$\chi ^{2}$ / degrees of freedom (d.o.f). Estimated EW limits
(right-most column, 90\% confidence) are calculated for fixed values of
$E =5.7$~keV and $\sigma =0.87$~keV. The name ``MOS" is used for
the MOS1 and MOS2 spectra modelled together; ``EPIC" - for both
MOS and pn spectra modelled together.}
 \label{tab2}
\end{table}

In Table \ref{tab3} we report the XMM-$\it{Newton}$ fluxes and
lensed luminosities together with the results from other authors,
obtained from {\it ROSAT} and $\it{Chandra}$ observations. The fluxes
obtained here are in good agreement with the $\it{Chandra}$ ones.

\begin{table}
\begin{tabular}{|l|c|c|c|} \hline
Satellite  & Energy  & $F,
10^{-13}$& $L, $ \\
Instr. & range, keV & erg/(cm$^{2}\cdot$ s)& $10^{45}$ erg/s\\
\hline
 XMM, MOS1  & 0.1-2.4    &   $2.5\pm$0.3                 &       \\
            & 0.4-8.0    &   $5.3_{ - 0.5}^{ + 0.4}$     & 11.8  \\
            & 0.2-10.0   &   $6.0_{ - 0.7}^{ + 0.4}$     & 13.5  \\ \hline
 XMM, MOS2  & 0.1-2.4    &   $2.6_{ - 0.4}^{ + 0.3}$     &       \\
            & 0.4-8.0    &   $5.4_{ - 0.6}^{ + 0.5}$     & 12.9  \\
            & 0.2-10.0   &   $6.0_{ - 0.7}^{ + 0.3}$     & 14.6  \\\hline
 XMM, pn    & 0.1-2.4    &   $2.4\pm$0.2                 &       \\
            & 0.4-8.0    &   $5.1_{ - 0.3}^{ + 0.4}$     & 10.4  \\
            & 0.2-10.0   &   $5.8_{ - 0.5}^{ + 0.3}$     & 13.3  \\\hline
 ROSAT      & 0.1-2.4    &   2.2                         & 4.2   \\\hline
 Chandra    & 0.4-8.0    &   $4.6_{ - 0.2}^{ + 0.4}$     & 10.0  \\
            & 0.4-8.0    &   $3.7_{ - 0.5}^{ + 0.6}$     &  8.3  \\\hline
 \end{tabular}
\vspace{0.05in} \caption{Absorbed power-law fitting of
XMM-$\it{Newton}$ Q2237+0305 spectra - fluxes and lensed
luminosities.} \label{tab3}
\end{table}

Commonly considered as evidence of an accretion disk in
the central engine of quasars or active galactic nuclei (AGN) 
is the detection of a
broadened Fe K$\alpha$ emission line at a rest-frame energy of
$\approx 6.4$ keV and a width in the range of 0.1--1.0 keV (see
Laor, \cite{laor}). Such lines were present in the spectra of 
42\% well-exposed objects from a sample of more than
100 AGN explored by Guainazzi et al. (\cite{guainazzi}) and in
the spectra of 2/3 of the 26 AGN from the sample analyzed by
Nandra et al. (\cite{nandra}). As shown by Reeves et al.
(\cite{RPT}) and Turner et al. (\cite{turner}), a broadened Fe
K$\alpha$ line can also be explained by the presence of an ionized
absorber with a column density of $\approx
10^{23}$~cm$^{-2}$. However, Nandra et al. (\cite{nandra}) show
that 1/3 of the cases (in the considered sample) cannot be adequately fitted with an ionized gas model, 
and for all of them the line broadening mechanism is well
explained by reflection from an accretion disk. In our case, the
apparent lack in Q2237+0305 of prominent absorption (mentioned
above) suggests that ionized absorber is likely not the case for this source.
Taking into account the redshift of the quasar, the line should be
observed between 2 and 3~keV. Dai et al. (\cite{dai}) found the
line at $5.7_{ - 0.3}^{ + 0.2}$~keV rest-frame energy, with
$0.87_{ - 0.15}^{ + 0.30}$~keV line width and $1200_{ - 200}^{ +
300}$~eV equivalent width (EW), relying on the $\it{Chandra}$
(2000--2001) data.

In the case of the XMM-{\it Newton} observation of Q2237+0305, 
the Fe K$\alpha$ line is not evident. We try to model this line with the
fixed rest-frame energy $E$ and line width $\sigma$ taken from
Dai et al. (\cite{dai}) to obtain an upper limit on its EW. The
results are shown in Table \ref{tab2}. On the other hand, if we
let the line energy and the width change over 
limited energy ranges (5--7~keV for the rest-frame energy and
0--1~keV for the line width), fitting of the pn spectrum yields
$E=6.0_{ - 1.0}^{ + 0.7}$~keV rest-frame line energy and
$\sigma<0.9$~keV for the line width; however, this result is not
statistically significant (according to the $\chi ^{2}$ test). 
For MOS1 and MOS2 such a treatment does not yield a meaningful result at 
all due to the lower counting statistics.

For completeness, we also fitted the pn spectrum of
Q2237+0305 with a model often used for the spectral 
modelling of AGN (Soldi et al., \cite
{soldi}, Schurch et al., \cite {schurch}): a Compton-reflected
(from neutral material) power-law energy distribution with
high-energy cut-off $E_{c}$ (pexrav, Magdziarz \& Zdziarski,
\cite{Magd}):

\[
\rho (E) = KE^{ - \Gamma }e^{ - \frac{E}{E_c }}.
\]

Neutral absorption in the lensing galaxy (zphabs) was also
included here. The model with reflection provides a slightly better
fit than a power-law ($\chi ^{2} = 1.05$ vs. $1.18$) which is,
however, not statistically significant. Search for the high-energy
cut-off ($E_c$) provided no results. The other parameters of this
fit are as follows: photon index $\Gamma=2.3\pm$0.3, relative
reflection parameter $R=2.3_{ - 1.2}^{ + 2.2}$, hydrogen column
density in the lensing galaxy $N_{H}=7_{ - 2}^{ + 3} \cdot
10^{20}$~cm$^{-2}$ and flux of $F=6.0_{ - 2.0}^{ + 0.2}
\cdot 10^{-13}$ erg/(cm$^{2}\cdot $s) in the 0.2--10~keV energy range.

\section{Lightcurves and variability}

The photon counts $Y_n$ in every $n$-th time bin represent
the total X-ray flux from all the images of GLS Q2237+0305
and have been obtained after background subtraction. The
time bin size in this treatment was varied from $T=0.1$ to 2~ks.

\begin{figure}
 \resizebox{\hsize } {!} { \includegraphics{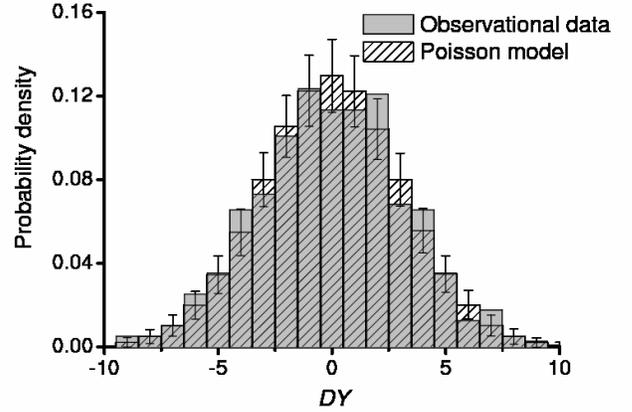} }
 \caption{The distribution of the count differences $DY_n = Y_{n+1} - Y_n$,
where $Y_n$ is the photon counts in every $n$-th bin. The observed
distribution is shown with the shaded histogram and the expected
distribution arising from Poisson noise is presented with the
stripped histogram. The error bars are at the $1\sigma$ level.}
 \label{fig6}
 \end{figure}

\begin{figure}
 \resizebox{\hsize } {!} { \includegraphics{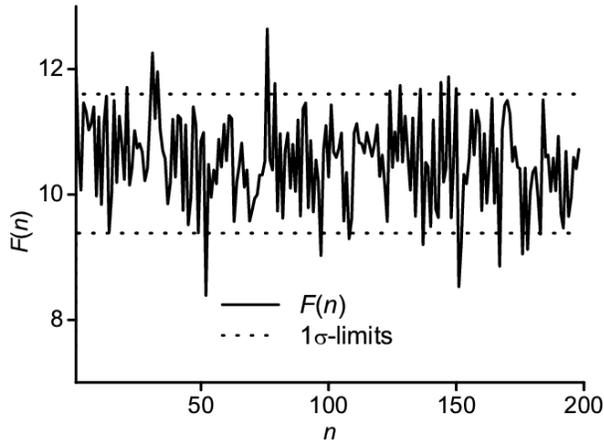} }
 \caption{Structure function (total data from all EPIC cameras). Dashed straight lines show upper and lower
 $1\sigma$-limits estimated for a Poisson model. The time bin size is 100~s.}
 \label{structure}
 \end{figure}

Variability in $Y_n$ (estimated by the ratio of standard deviation
$S$ to the mean $M$)  is not  noticeable in the lightcurves (Figs.
\ref{fig2}, \ref{fig3}, \ref{fig4}). Numerical estimates are
presented for 0.5~ks and 1~ks bin sizes (Table \ref{variability}).
Note that we do not observe a significant correlation (less than
0.2) between pn, MOS1, MOS2 data for various bin sizes which
indicates that variations of the photon counts are simply due to
their Poisson dispersion. We examined this question in more detail
on the basis of the model of stationary Poisson process for the
photon count numbers (i.e., with no variability). We
considered the statistics of the observed count differences
$DY_n=Y_{n+1}-Y_n$ and compared them to the statistics of
simulated count differences in the case where the differences
follow Poisson statistics. We find that the distribution $DY$
obtained from the observational data is very close to our
simulated $DY$ distribution that assumes a Poisson model for the
pn, MOS1 and MOS2 cameras. The distribution of $DY$ from the
combined data of all the EPIC cameras is shown in Fig.
\ref{fig6}.  The $\chi^2$ test yields 91\% probability that the
distribution in Fig.\ref{fig6} is a realization of a Poisson
process. Also we present in Fig.\ref{structure} the structure
function
\[
F(n) = \frac{1}{N}\sum\limits_{i = 1}^N {(Y_{i + n} - Y_i )^2},
\]
which agrees with Poisson model.
Therefore, the hypothesis of constant X-ray flux from the
source is compatible with the experimental data. The estimates 
reported above also apply for the 5--10~keV energy interval, 
with no substantial changes in variability.

\begin{table}
\begin{tabular}{|l|c|c|c|c|} \hline
 Camera  & $T, $    &   $V,$   & $\nu$ \\
         & ks   &     &   \\
 \hline
  pn   & 1 & $\,0.13$  &0.67\\
  \hline
  pn   & 0.5 & 0.24 &0.89\\
  \hline
MOS1 & 1   & 0.25 &0.87  \\
 \hline
 MOS1 &  0.5 & 0.40  &0.99 \\
 \hline
 MOS2   & 1&0.33 &1.13\\
\hline
MOS2   & 0.5  &0.40 &0.96\\
\hline

 \end{tabular}
\vspace{0.05in}
 \caption{Parameters of the photon counts: $T$ is the time bin
 size. Variability is characterized by $V=S/M$,  $S$   is the standard deviation and $M$ is the mean of
 $Y_n$; the parameter
  $\nu=S/\sqrt M$ compares $S$ with the calculated Poisson variance. 
The correlation of the net signal
 with the background in all the cases is less than $0.2$.}
 \label{variability}
\end{table}

\section{Discussion }

We analysed the  XMM-$\it{Newton}$ data (May, 2002) on the total
X-ray flux from all four images of the GLS Q2237+0305 ``Einstein
Cross". The flux value in the 0.2--10~keV range is
$5.9\cdot 10^{-13}$ erg/(cm$^2\cdot$ sec), corresponding to
a lensed luminosity of $1.25\cdot 10^{46}$ erg/sec. In
order to obtain the true quasar luminosity, this value must be
corrected, taking into account the lens magnification. The most
recent model estimate of the magnification is $16^{+5}_{-4}$ (Schmidt et al., \cite{schmidt}). 
Fitting the X-ray spectra in the
0.2--10~keV energy range by an absorbed power-law yields the
photon index $\Gamma=1.88\pm0.10$ and hydrogen column density in
the lensing galaxy $N_H=(7\pm$2)$\cdot 10^{20}$~cm$^{-2}$. This is
in satisfactory agreement with the results of Dai et al.
(\cite{dai}).

The XMM-$\it{Newton}$ lightcurves obtained after the
background subtraction  are compatible with the hypothesis of a
constant flux from the source. An additional
argument in favor of no variability is the absence of a
significant correlation between the light-curves from different
cameras. This does not completely rule out source variability: 
(i) it may be obscured by the Poisson variance; (ii) it may be
smeared out in the total X-ray flux from all the images that have
different time delays; (iii) there may be periods of different
variability amplitudes. These periods may correspond to
the lifetimes of X-ray emitting blobs orbiting the black hole and
therefore we might expect variability periods of the order of the
revolution time around the black hole. In any case, the net
variability during the XMM-$\it{Newton}$ observation cannot
be large. As for concern (iii), we note that the {\it ROSAT} 
(1997) observations  did not reveal significant variability. On
the other hand, the variability was considerable during the 
$\it{Chandra}$ (2000, 2001) observations (Dai et al., \cite{dai}).
This suggests the presence of slow variability changes with
characteristic times much longer than the inner variability
timescale $\tau_v\sim 1$~hour following from the results of Dai et
al. (\cite{dai}).

Light-travel arguments imply that the one hour time-scale
variability corresponds to a size-scale of the X-ray emitting
region responsible for this variation of $R_X\sim\tau_v
c/(1+z_s)=4\cdot10^{13}$ cm. An interesting question is whether
$R_X$ represents an inhomogeneity in the accretion disk or the
entire X-ray emitting region. This question might be answered,
provided that the black hole mass in QSO 2237+0305 is known.
Kochanek (\cite{kochan}) estimated this mass, implying a
radius of the innermost stable circular orbit of $3r_g\approx
(4\div 22.5)\cdot10^{14}$ cm  (see, e.g., Chandrasekhar,
\cite{chandrasekhar}) in the case of a non-rotating black hole
(where $r_g=2GM/c^2$ is the gravitational radius). The inner
radius of a real accretion disk must be even larger. However, the
mass estimate of Kochanek (\cite{kochan}) is model dependent and
it is not sufficient to make a rigorous judgement concerning a
comparison with the order-of-magnitude estimate of $R_X$.

One may ask how this inhomogeneous structure (if it exists) would appear 
in the presence of microlensing. The theory of
high-amplification events in the case of a small
continuous source is well known (see, e.g., Grieger et al.,
\cite{grieger}; Shalyapin et al., \cite{shalyapin}; Popovic et
al., \cite{popovic} and references in these papers). In
the case of one emitting spot, the microlensing
amplification scales as $\sim R_X^{-1/2}$. Obviously, in the case of a
large number of small inhomogeneities, the total amplification may
be smeared out over the source. However, different regions of the
time-dependent inhomogeneous structure will be differently
amplified, resulting in an increase of the variability amplitude in
microlensed image. Note that independent contributions of this
microlensing effect may reduce the correlation between the lightcurves
in different images. On the other hand, any prominent
peak in the lightcurve without repetitions after the expected time
delays can be considered as a probable sign of microlensing.

The idea of inhomogeneity of the quasar core in GLS has been
repeatedly discussed from different viewpoints (see, e.g., Wyithe
\& Turner, \cite{wyither1}; Wyithe \& Loeb, \cite{wyither2};
Schechter et al., \cite{schechter}). Some evidence in favor of
inhomogeneous structure may be found in optical observations of
different GLSs, where the rapid flux variations are observed on
the background of slower changes (Colley \& Schild, \cite{colley};
Paraficz et al., \cite{paraficz}). Note that there is an
alternative explanation for these variations that invoke a
planetary-mass object in the lensing galaxy (Schild \& Vakulik,
\cite{vakshild}, Colley \& Schild, \cite{colley}). 
It might be possible to distinguish between these two models by long-term 
optical and X-ray monitoring of microlensing events in AGN.

Observations of the resolved X-ray images with $\it{Chandra}$ are more informative for
the Q2237+0305 variability studies and determination of the time
delays. Nevertheless, observations of the total X-ray flux with
XMM-$\it{Newton}$ may provide valuable complementary information
on this issue. In the case of XMM-$\it{Newton}$, 
sufficiently long observations are needed to obtain a good 
result for the autocorrelation function. To avoid uncorrelated
contributions from different images to the
autocorrelation function, the monitoring period must be
larger than the differential time delays in
this system. In the case of Q2237+0305, the available measurement
interval of almost 42~ks is clearly not long enough. The model
prediction (Schmidt et al., \cite{schmidt}) of maximal delay for
Q2237+0305 is 17~hours. Large delays ($>35$~ks) are also predicted
by earlier models of Q2237+0305 (Rix et al., \cite{rix}; Schneider
et al., \cite{schneidtau}). This is consistent with the analysis
of the optical lightcurves of Q2237+0305 (Vakulik et al.,
\cite{vakulik}), which yields an upper limit for the delays of
about $\sim 2$~days.

While waiting for new observations one may  ask whether in
principle it is possible to determine the variable X-ray signal
$x(t)$ from the source if the total flux from all the GLS images
is available. Typically, this problem has an infinite number of
solutions. However, if we know $x(t)$ over a sufficiently large
initial interval, then  $x(t)$ may be recovered uniquely from 
the total flux lightcurve (Zhdanov et al., \cite{zh}). 
This may be of interest if one needs to combine
results of long GLS monitoring of the total flux (cf.
XMM-$\it{Newton}$) with shorter observations of separated
different X-ray images (cf. $\it{Chandra}$).

\begin{acknowledgements}
We are grateful to the referee for very helpful comments and suggestions.  
E.V.F. and V.I.Z. thank the VIRGO.UA center in Kiev, 
IASF--INAF and Astronomy Department of Bologna University for offering 
informational and technical facilities and financial support. 
C.V. acknowledges partial support from the
ASI--INAF grants I/023/05/0 and ASI I/088/06/0. 
E.V.F. is also grateful to Dmitry A. Iackubovsky and Pawel Lachowicz 
for their guidance in using the XMM-{\it Newton} software. 
\end{acknowledgements}

\end{document}